\documentclass{article} \begin{document}

\title{The Epistemic Landscape: a Computability Perspective}

\author{Fr\'ed\'eric Prost\\
	            Laboratoire d'Informatique de Grenoble, Univ. Grenoble Alpes, CNRS,\\
	    Grenoble, France \\
	    {\tt frederic.prost@univ-grenoble-alpes.fr}}

\date{\today}

\maketitle

\begin{abstract}
   By nature, transmissible human knowledge is enumerable: every sentence, movie, audio record can be encoded in a sufficiently long string of 0's and 1's. The works of G\"odel, Turing and others showed that there are inherent limits and properties associated with the fact that language technology is enumerable. G\"odel's numbering technique is universal for enumerable structures and shows strong limits of the language technology. Computability theory is a particular example: programs can be numbered and all sorts of limits can be studied from there. Computability is also at the heart of science since any experimental validation of a theory supposes that theoretical results have been computed, then checked against concrete experiments. It implies that limitations on what is computable ultimately are also limits of what we understand as "scientific theory", and more generally to all the transmissible knowledge. We argue that it is fruitful to look a epistemology from a computability perspective. We show that it allows to precisely define different kinds of knowledge acquisition techniques, and helps the study of how they are related to one another. 
\end{abstract}

%%%%%%%%%%%%%%%%%%%%%%%%%%%%%%%%%%%%%%%%%%%%%%%%%%%%%%%%%%%%%%%%%%%%%%%%%%%%%%
%%%%%%%%%%%%%%%%%%%%%%%%%%%%%%%%%%%%%%%%%%%%%%%%%%%%%%%%%%%%%%%%%%%%%%%%%%%%%%
\section{Introduction}
\label{sec:Introduction}
All of the transmissible human knowledge can be included into computer science. Indeed, every book, every recorded teaching class or every picture can be digitized and put inside a computer. One can even advocate that all of the human literature is just made of different outputs of a single software: namely a word processor to which one has provided the right inputs (the letters and typesetting information to print down the book). This remark is more than a cheap trick. It is in essence G\"odel's technical point to prove the incompleteness of the arithmetic. Indeed, the first step of the G\"odel's proof is to encode mathematics within themselves. Propositions and proofs are encoded into numbers. The fact that some sentence is the proof of another sentence can then be viewed as a peculiar relation between two numbers. The final step of G\"odel's proof is to encode a version of the liar's paradox. It is done by formulating a proposition stating something like: "I am not provable". The only way for the system to hold is to have true sentences that are not provable. This result, that truth does not necessarily implies provability, is paramount from an epistemological point of view. Turing made a mechanical version of this theorem: the undecidability of the halting problem. Turing exhibited a peculiar machine, the universal Turing Machine, that is able to simulate all Turing machines. This allows him to build a self-referential paradox within this context: a machine that halts if and only if it doesn't halt. 

\medskip
The fact that all human knowledge can be seen as a part of computer science shows that computability theory has a special status from an epistemological point of view. Moreover, there are strong results on what is computable and what is not computable. These limits do apply to all fields of the human wisdom, not only computers. It turns out that looking at the epistemological landscape from this point of view is fruitful and sheds new lights on the relations between different categories of knowledge. In this context Turing machines are not seen as mere computer but as an abstraction of an unequivocal computation, or reasoning, that can be explicitly described.

\subsection*{You cannot explain everything}

Whatever one may think of physical reality the only way to talk about it is in a countable way. Indeed, every statement one can ever make about the world are sentences written with letters drawn from a discrete alphabet. The same argument can be said regarding measurements: at the end of the day one has to record the measurements using words. The fact that the results of an experiment is a video does not change that: the sentence associated with the video is just the huge binary number composed by the bits of the file recording the video. This is a casual version of G\"odel numbering technique. It is possible from there to use the diagonal argument to show that the hope of a theory of everything is not possible without having to unpack precisely what is intended behind words such as {\em explanation} or {\em phenomenon}. What is only needed are the two following basic requirements:
\begin{itemize}
   \item An explanation is a sentence linked to a phenomenon of interest. It can be instantiated. For instance the classical Newtonian mechanics gives an explanation for the motion of objects. But for each system considered you have a different instantiations of this explanation. Explaining why a shell $A$ will land on a particular location $X$ is not the same explanation as why shell $B$ will land on location $Y$. Those are a two separate explanations even if they originate from a single theory.
   \item An explanation is a phenomenon. Indeed, an explanation is a text that is written trying to make sense of something. In itself it is a phenomenon, it has an expression in the real world (the piece of paper on which it is written for instance, or the trace of the modification of a hard drive where the text is recorded).
\end{itemize}	   

It would be very hard to understand what the term {\em explanation} means without the first point. The second point is maybe more debatable but appears very compelling: any discourse is a phenomenon of the world, and as such, one would hardly not take them as phenomenon themselves.

Those two ingredients are enough for the self-referential paradox machine to work. We can follow the path of Cantor, G\"odel and Turing to show that it entails that, mathematically, it is not possible to have an explanation for every phenomenon.

Since explanations are enumerable we can order them (using the alphabetical order for instance). It means that there is a function “Explain” from integers to sentences such that Explain(i) is the i-th explanation.
Now, one may consider questions like: "are there any links between explanation $i$ and explanation $j$?". Any explanation for such questions are phenomenon. But it entails that the size of all phenomenon is at least equipotent to the power set of natural numbers. For that, one has just to consider a sligthly different kind of questions: "is explanation $i$ related to any explanation in the set $S$?", where $S$ is any subset of natural numbers. A direct application of Cantor's diagonal argument concludes because the set of natural numbers is not equipotent to the set of parties of natural numbers. Therefore, there are (infinitely many) phenomenon that can't have an explanation. 

\subsection*{Negation and provability}

Of particular interest is the role played by negation. One of the important result outlined by the works of G\"odel and Turing is that negation does not commute with provability. In other words "knowing that no" is not equivalent of "not knowing". The halting problem highlights this remark.  When a machine is trying to prove something and does not halt, one cannot say anything about the result: it does not exist. It requires a computing machine to have this idea of "not having an answer". Indeed, in mathematics partial functions are completed by the addition of a bottom element. This bottom element is not in the codomain of the function and denotes that the function is not defined on particular points. But doing so is cheating because something is known that shouldn't. In a trivial way it is the difference between {\em known unknowns} and {\em unknown unknowns}. On the other hand, if the result of the function is seen as the result of a computation done by a machine, then the only way to have a truly undefined answer is to not have an answer at all. Hence the crucial importance of the halting problem. 

This technical result, the fact that negation and provability do not commute, has numerous repercussions in human knowledge. One example, is the judiciary version of {\em truth}. By default the burden of proof is on the one who declares not on the one who denies. A falsifiabilty definition of the scientific method, as proposed by Popper, is also an illustration of this point (see \ref{sec:Popper} for a more elaborated discussion on this point). On the other hand, the precautionary principle, is its opposite. Indeed, a central point of the precautionary principle is that the doubt supersedes scientific knowledge. In all rigor, one may always pretend that a danger will occur sometime: you can test the system $x$ times (days, years, generations, whatever), even if each test is a success one can always pretend that problems will occur at the occurrence $x+1$. There is no end to this process that is ultimately semi-decidable: if there is a problem, one day you will find out, otherwise you are going to search for potential problems ad infinitum. But it is possible that this principle is of use when you deal with very important issues involving important repercussions \cite{Taleb_2014}.  

\medskip
In this paper we advocate that epistemological categories can be defined from a computability point of view. In the same way that some problems are "more" undecidable than others (see the arithmetical hierarchy \cite{Rogers1987}). The degree of confidence in the acquired knowledge goes from mathematical certainty to fuzzier ideas. Linked to this is the problem of continuity between ideas and reality. The basic problem is that knowledge should at least be a safe approximation of the reality. The map should faithfully represents the territory even if is necessarily an abstraction of it. Our aim is not to discard non-scientific knowledge but rather to acknowledge that human knowledge is inherently heterogeneous. Notably, there is this apparent paradox: there is a scientific proof that not every part of human knowledge can be scientifically established.  

A tricky part is that some pieces of reality are subjective. Think pain, emotions etc. It leads to a kind of Heisenberg uncertainty principle in humanities. Indeed, by their very definitions the object of study is subjective. It leads to overly context dependent truths. Adding to that a layer of complexity when emergent properties of interacting subjects are considered, and one soon finds that social sciences are by no way comparable to hard sciences. This last statement can be shown as a technical result.    

\medskip
In some sense this essay, from a broad perspective, is an interpretation of the Church-Turing thesis. Our take on it is that it is not that it is physically impossible to have a machine more powerful than a computer, but that it is impossible to build one. It points a limit of the formalization efforts based on the language technology. The point being that one cannot {\em build} a machine if one doesn't have a set of instructions to do it. But the plan (which is in fact a theoretical formal system describing your machine) cannot leads to something more than a Turing machine. Indeed, as soon as an alphabet is used in order to describe a piece of knowledge, it falls prey to computability limits. But there is no reason altogether to think that this limit is a limit of nature itself. Besides, even if no one knows how to build an artificial general intelligence, it is very easy to produce one: one only needs to have kids. The point being that it is not because one doesn't know how something works that one can't do it. But it would be rather peculiar to present this as the result of an application of structured knowledge.

\subsection*{Outline of the paper}
Different tiers of knowledge building strategies are going to be investigated. I follow an approach based on computability theory. The idea is not to build a formal morphism between the different levels of the arithmetical hierarchy and the different way one can gather knowledge, "à la" Curry-Hoard \cite{Girard_1989},  but rather to define a taxonomy of how knowledge can be built. There are several tiers, for each tier there are different notions of proof, falsification and certainty in the production of knowledge. The first tier of knowledge amounts to decidable problems, and is basically represented by mathematics, it is discussed in section \ref{sec:Theory}. The second tier, tackled with in section \ref{sec:Popper} is the scientific approach: knowledge belonging to this tier are such that there should be a workable way to negate them, that is a constructive way to find counter-examples (which is enumerability of disproof). The third tier is enumerability for proofs and is studied in section \ref{sec:Grey}: knowledge that are adressed in this section are such that either one day they will be prooved, or they are not provable at all. In the fourth tier, section \ref{sec:Emergent}, undecidable problems are tackled. It happens when nor proof, nor proof or absence of proof can be achieved. Though one may still builds knowledge on shakier grounds. Typically, functionnal knowledge that emerge from repeated confrontations with reality. This kind of knowledge is not scientific, but cannot be discarded at once. The only mechanism to uncover this kind of knowledge is through survival: if a rule of thumb is old and has worked reasonnably well for long period of time, one may have a certain level on confidence about it. Then, in section \ref{sec:Descriptive}, I adress another tier: descriptive knowledge. It is the dual tier of the mathematical one. It is knowledge that comes from reality and does not bring anything more than itself. It has the static aspect of mathematics but is finite by nature. It is only interesting because of its direct mapping to reality. For instance one may describe a country with a map, but this map cannot be used to infer any new knowledge like theories and model do. In section \ref{sec:Fiat} is discussed a tier in which the objects of study itself are subjective: humanities. Then, in section \ref{sec:artsorscience} deals with the particular bodies of knowledge which are hard to sort because they cover wide areas. Due to scaling difficulties they appear both as scientific at some scales and not scientific at other scales. Typical example are economics and medicine.  Finally, I conclude in section \ref{sec:Conclusion}.

%%%%%%%%%%%%%%%%%%%%%%%%%%%%%%%%%%%%%%%%%%%%%%%%%%%%%%%%%%%%%%%%%%%%%%%%%%%%%%
%%%%%%%%%%%%%%%%%%%%%%%%%%%%%%%%%%%%%%%%%%%%%%%%%%%%%%%%%%%%%%%%%%%%%%%%%%%%%%
\section{The theoretical area: the world of abstractions}
\label{sec:Theory}

The mathematical world seems rigid at first sight: things are defined once for all. The whole universe of the discourse is contained within the limits of a small and precise set of definitions. Mathematical ideas have a very Platonic flavor of eternal truths floating around us. In the same way that all legal chess games are contained in the definition of the rules of the game. We can only explore a limited area (play so many chess games) of mathematics due to the finiteness of our existence. But, in principle, truth remains eternal and outside the reach of anything linked to reality: once the axioms and deduction rules are properly defined, the only work left to do is to uncover truths that were already implicit in the foundations. One can go even further and advocates that those mathematical objects exist outside human existence. In this mathematical ideal world, proofs are static objects obeying syntactical rules: one can make a computer to check whether or not the proof is correct. For instance proof-checkers like Isabelle \cite{Nipkow}, Coq \cite{coq} etc. are concrete examples of this.   

Actually modern approaches of mathematical proofs are more complex than that: game semantics \cite{Majer09} or ludics \cite{Girard01} are examples of interactive proof frameworks for which there is not this idea of eternal truths. Proofs are seen as Socratic dialogue between two players, one player trying to convince the other of the truth of a statement. But even without considering those frameworks, things are not as fixed as one may think:

\begin{itemize} 
   \item Firstly, one of the major work of mathematicians is not to find proofs, but rather to define useful notions and structures. Mathematics is an ever expanding domain where new useful abstractions are designed every day. What does "useful" means in this context? This question is very deep and I will not tackle it more than by taking a functional point of view. An abstraction is useful if it helps you to achieve a task in the real world. A trivial example is given by the concept of numbers which is arguably one of the oldest abstraction (with painting which is nothing more than an elaboration on shadow, aka projection from a mathematical perspective). The number abstraction allows to keep track of the herd for instance. Addition, multiplication, equations and so on naturally derive from the first abstraction and find other applications. Similarly, the basic notions of geometry: points, lines, planes etc. are abstractions (pushed to the limit) of shapes. The fact that those abstractions amount to the same one, as brilliantly showed by Descartes in \cite{Descartes}, can appear as unreasonably surprising (to borrow from \cite{Wigner}). But it is not relevant from our epistemological perspective. 

   \item Secondly, even if we stick to a static point of view that once the definitions have been laid out, and one would only has to "uncover" truths supposedly existing from the definition, there is still work to do to produce the proofs. This work is not free and has to be seen as a positive increase of knowledge. Let us consider again an analogy on games: even if it is well-known that in the game of Hex the starting side is theoretically winning (using a nice argument of strategy-stealing by Nash see \cite{Beck}), it does not give a clue on how to play an actual game. 
\end{itemize}

The role of computability looks rather small in such an area in which everything seems to follow logically, one could even say mechanically, from the definitions. Indeed, even if mathematicians have to work to produce a proof, it would appear at first sight that this work could be done by a machine displaying all the proofs in a way similar to the one that an infinite number of monkeys randomly hitting keys of a typewriter would end up writing all of Shakespeare's complete works. It turns out that things are more complex than that. Indeed most theorems are not provable. This striking and famous result is due to  G\"odel, and known as the incompleteness theorem. There are several version of it: the halting problem (writing a program that can decide whether or not a program will eventually halt) may be seen as a particular instance. Therefore, the infinite monkey remark does not apply. Indeed if it turns out that the theorem for which you are looking the proof belong to the list of truths not provable. Thus, a proof is nowhere to be found.

At the end of the day the certainty in knowledge acquired in the mathematical area is built-in. Mathematical proofs, when they exist, stand on their own: there are no experimental ways to falsify (or justify for that matter) a mathematical result. The contact with reality only occurs via the concrete uses of the mathematical theories. But, by themselves, those mathematical theories are separated from real life. It is possible though to misapply mathematics, that is to model a real life phenomenon and find contradictory results. In this case the contradiction found is in the modeling process not about the mathematical result in itself. This separation from reality comes with a cost: the one of consistency. Ultimately, the only thing asked to a mathematical theory is to not allow both $A$ and not $A$ to be true at the same time. This is the only guardrail from erroneous judgments.

All in all the mathematical knowledge has a very platonic flavor in the sense of the theory of forms: in the mathematical area truths look eternal, without having to be located somewhere, etc. To paraphrase the quote from \cite{Moore2008}: “Mathematics are realityproof”. This quote works somehow both ways. In mathematics you can go as far as your imagination will lead you with the only caveat that you have to have consistency.

%%%%%%%%%%%%%%%%%%%%%%%%%%%%%%%%%%%%%%%%%%%%%%%%%%%%%%%%%%%%%%%%%%%%%%%%%%%%%%
%%%%%%%%%%%%%%%%%%%%%%%%%%%%%%%%%%%%%%%%%%%%%%%%%%%%%%%%%%%%%%%%%%%%%%%%%%%%%%
\section{The scientific/rational area: Popper}
\label{sec:Popper}

There is a vast consensus on what constitutes the scientific method in order to grasp a phenomenon.An hypothesis is defined, which amounts to a formal system, based on your observations. Then, experiments trying to disprove the hypothesis are elaborated and tested. It is more or less the philosophy of the falsifiability proposed by Popper \cite{Popper59}. The problem with this description is that it does not ellaborate on what "emitting an hypothesis", and using this hypothesis in order to make predictions, are. It hides the fact that computations have to be actually be made within the theory to compare the actual measurements/experiments with what the theory forecasts. The very fact of forecasting implies that the scientific method is subjected to the Church-Turing thesis. Because at the end of the day one can only compute what is computable... by definition. And if all what is computable is well captured by Turing machines, then the scientific area cannot be beyond that. Hence the scientific method can only be used for formal systems in which predictions are computable. Computability limits do indicate hard limits of what the scientific method is subjected to.

\medskip
If the existence of a limit for the scientific method is known, by definition it cannot be defined precisely. Indeed, drawing that frontier would imply that the halting problem would be solved by being able to tell which functions are computable from which are not computable. The scientific quest is a never ending one. 

On one hand, one will always try to build more accurate theories of reality, but on the other hand one may never know for sure when this quest becomes illusory. A concrete example given by \cite{CubPerWo15}. This work shows, that we can, by design, build systems that, following our current formalism to understand the world, behave in an uncomputable way. The idea is to build a structure that has some physical properties if and only if some Turing machine halts. It is conceivable, though, that a new theory can be designed in such a way that this problem becomes computable. But first, in this new theory there will be new uncomputable phenomenon (this is not without reminding the building of the arithmetical hierarchy in which new unsolvable problems are built on top of oracles solving previously unsolvable problems). And second, we can't know for sure that such a new theory is possible before having found it. It is very similar to the task of finding decidable subclasses of an undecidable problem.

\medskip
The falsifiability requirement of the scientific method means that science is only considered established for phenomenon for which it is possible to provide a constructive way to find counter-examples. By constructive it is meant that the experiments should be practically implantable. Not “in principle” implantable, but in real life with a constructive way to build the experiments.

From a computability point of view the scientific/rational claims can be seen as the class of theories for phenomenon for which counter-examples search is decidable. It means that there exists a way to decide, in a finite time, whether or not there are counter-examples. It is not always like that. For instance one way to prove the Goldbach’s conjecture (every integer greater than 2 is the sum of two prime numbers) would be to prove that there is no counter-example, which can be shown equivalent to proving that the research of a counter-example for this conjecture will never halts. Indeed, to find a counter-example one has to try all even numbers one by one, but this search (that can be seen as particular computer program) will never end if it turns out that the conjecture is true.

Certainty is achieved, in the scientific/rational area, within a global context: a scientific proof implicitely supposes that Nature does not change the rule of the game across time. More than that, it supposes that such a rule exists in the first place. In fact, science can only reliably talk about phenomenon that have this property. So the notion of proof is rather different in this area than in mathematics: actually what is called a proof in the scientific area is an absence of disproof with relation to a specific context. From a computability point of view it is the class of phenomenon for which counter-examples search is decidable.

%%%%%%%%%%%%%%%%%%%%%%%%%%%%%%%%%%%%%%%%%%%%%%%%%%%%%%%%%%%%%%%%%%%%%%%%%%%%%%
%%%%%%%%%%%%%%%%%%%%%%%%%%%%%%%%%%%%%%%%%%%%%%%%%%%%%%%%%%%%%%%%%%%%%%%%%%%%%%
\section{The Semi-Scientific area: Science either for the answer or the absence of scientific answer.}
\label{sec:Grey}

Refutability is a useful safeguard: it ties abstract theories with the real world in a practical way. But there are pieces of knowledge that are not prone to refutation while not being useless speculations. Those pieces of knowledge are charactarized by an absence of positive evidence, but for which one never finds a counter-example. The semi-scientific area is the dual of the faillability requirement in the following sense: f the claim is true, then a constructive proof will be found, but if the claim is false, then we can prove that no proof that the claim is false is possible. 

Some issues inherently fall in this category. I may refer to them as "semi-scientific" knowledge (reminding semi-decidability). It means that for them either, one day, a scientific answer can be settled, or it is possible to scientifically assess that noone will ever have a scientific answer. It differs from the scientific knowledge in the fact that there is no practical way to falsify them, but there is, in principle, a way to validate them.  It is an equivalent Goldbach's conjecture mentioned in section \ref{sec:Popper}. No proof has been found that this conjecture is correct, and no counter-example neither, which would be a falsification of this claim. If this conjecture is true but not provable in ZFC this situation may continue forever (assuming that the independance from ZFC axioms cannot be proved too). Therefore a statement like "Goldbach's conjecture is true or we will never find a proof that it is false" would be (assuming that it is not provable in ZFC) correct and not refutable at the same time.

A typical example is the following statement: Is a phenomenon like conscioussness Turing-computable? 
\begin{itemize}
  \item If it is Turing-computable, hence someone may finally settle the question by bringing out an algorithm that simulates consciousness, whatever is understood by the term "consciousness". It is not difficult to imagine that experiments could easily been set-up once a working definition of "consciousness" has been layed out. 
  \item If it is not Turing-computable, it means that it is not possible to describe this phenomenon using a formalism. Indeed, if it were the case one would be able to make a theory of concioussness, this theory could be written down into a computer program which would contradict the fact that this phenomenon is not Turing-computable.  
\end{itemize}
It is a peculiar kind of knowledge: either it will become scientific (but without any certainties until a working algorithm has been produced), or it is  sure that the answer, from a scientific point of view, can never be established. 

\subsection*{Provability and Refutation}

What is witnessed here is this absence of commutativity between provability and negation. The scientific area of knowledge where there is decidability of counter-examples search. It means that if a counter-examples exists there are decidable procedures to produce them. The semi-scientific area of knowledge has semi-decidability of positive instances. It means that if the proposition is true, one day a constructive proof will be found. But if the proposition is not true, it is established that a scientific proof of that fact can never be established. Relatively with the scientific area there is no decidability on how to produce counter-examples in this area. 

In this area though, there is the hope of a scientific proof. It makes this tier of knowledge distinct from the tier examined in section \ref{sec:Fiat} in which a scientific cannot even be hoped for. 

%%%%%%%%%%%%%%%%%%%%%%%%%%%%%%%%%%%%%%%%%%%%%%%%%%%%%%%%%%%%%%%%%%%%%%%%%%%%%%
%%%%%%%%%%%%%%%%%%%%%%%%%%%%%%%%%%%%%%%%%%%%%%%%%%%%%%%%%%%%%%%%%%%%%%%%%%%%%%
\section{The Emergent area: Functional truths and the test of time}
\label{sec:Emergent}

Are the scientific, or the hope of a scientific, answer, the only ways to approach knowledge acquisition? It turns out that the answer is no. There is another technique based on evolution/selection, to expand knowledge: it is the knowledge that works in real world and has passed the test of time while not having a scientific explanation (ie a working theory in which the phenomenon can be dealt with from an abstract point of view). The typical examples are myths and religions. One way of seeing these functional truths is that they are collections of stories that have been transmitted through generations: useful ones have survived, unuseful ones have disappeared. Moral values, indications on what is, and how have, a good life are example of such truths. They are typically ideas that go backward from a causality point of view. Causality is the attempt to describe the present as a function of the past. Moral values, political decisions, etc., go the other way around: in some sense they are self-fulfilling prophecies. It is because you want some future event to take place that you modify the present by adpoting certain behaviors. In some sense it looks like that it is the future that is the root cause for the present.   

This tier of knowledge is very slippery because there is nor internal (coherence), nor external (reality check in the form of properly defined experiences), surveillance mechanisms working as a guardingrail. The worst part being that, ultimately, it cannot even be proved that a particular kind of knowledge belong to this tier (contrary to knowledge discussed in section \ref{sec:Grey}). From a computability point of view it is the equivalent of the Post's theorem: if two complementary sets are semi-decidable, then each set is in fact decidable. 

In particular it means that one cannot hope to make a theory out of them. And even if, at some point, an explanation is found, there is no warranty that this is the "right explanation". A concrete example is given by the jewish food law forbidding the simultaneous consumption of milk and meat. Nowadays it is knowns that calcium makes the absorption of iron more difficult \cite{Lnnerdal}. Moreover, due to the endemic presence of Malaria in this part of mediterrannea, there were selective pressures leading to a higher prevalence of thalassemia than average. Indeed, thalassemia offers some kind of protection against Malaria \cite{Haldane}. But on the other hand people having thalassemia suffers from anemia and have iron deficiencies. This theory is maybe a valid explanation of this old jewish diet rule. But other explanations are also valid: maybe it was just a rule to enforce ingroup/outgroup distinction, or maybe because a large part of this population is lactose intolerant and it makes digestion easier for them. Maybe it is a mix of those explanations or something else we don't have any clue about. The point is: we will never know for sure why this rule has been edicted and preserved. Moreover, there is no hope to have a rational explanation meeting the standard of a scientific proof one day. Because, at the end of the day, it is a law that has been layed down by human beings, not a law of nature.

\subsection*{Functional truths are not subjective}
Another point is that one cannot build those type of knowledge on his own. Because by construction they are ideas that need to be tested over very long period of times to prove their utility. Typically it takes many generations to consider that an idea of this flavor is not just a subjective random idea. So even if one profers such an idea, one will never know, or have a proof, of its efficiency. Such knowledge actually are a product of the interactions between humans. They are ideas vetted by adoption rate and adequation to the context in which they are used. Other typical instances are human institutions: their validity may change over time, they evolve slowly without one being able to precisely pinpoints particular instant and circumstances at the origin of the evolutions. The typical example is the one of natural languages.

\subsection*{Undecidability and relativity in practise}

This area of knowledge is peculiar since it does not rely on consistency. It is sometimes used as an argument in the discussion around religions vs atheism. The point being that religions should not be relevant, because they contradict each other. Let us reconsider the example of the jewish dietary rule regarding meat and milk. It could be the case that it is simply for environmental reasons that it is a useful rule. Therefore, there is no contradiction if in another region of the earth, where then environmental context is not the same, that other dietary rules are edicted more in accordance with local peculiarities. Actually, what would appear as suspicious would be that religious beliefs were homegeneous around the world, and across time.

On one hand, from a computability perspective, functional truths may correspond to undecidable phenomenon. It looks like the Goldbach's conjecture, but from a practical/real point of view. One cannot find meaningful counter examples but yet cannot exhibit a theory of why such rule is correct. On the other hand, from a historical/sociological point of view, functional truths appear as context dependent laws that may only be valid in some specific contexts.

The fact that such kind of knowledge takes the form of religious revelation is not surprising: since there are no real "explanations", they have to be taken for granted. This idea is well captured by the saying "God works in mysterious ways" which is not as sarcastic as one would may think at first, but as a reflection of undecidability/incompletness phenomenon in everyday life. Religious beliefs could be seen as a method to render a partial function complete with the addition of a bottom element: here the divine explanation fills the gap for all unexplainable phenomenon.

\subsection*{Time as the only guardingrail}

This land of knowledge brings the risk of misunderstandings, because of the absence of guarding rails like consistency in mathematics or falsifiability with the scientific method. Attempts to build theories (which ultimately may not possible to either to prove or disprove such kind of knowledge) may spiral out of control leading to religious wars since there are no established ways to validate/invalidate this kind of knowledge. The only practical way lies in the test of time which, admittedly, is a loose definition (how long is "enough"?). It is possible nowadays to somehow use the ease to widely spread information to see if an idea is "correct" more quickly. If the idea is largely used (and therefore tested in a multitude of contexts). But one must always take into account that time is required in order to cope with phenomenon having a very low probability of happening while having a huge significance.  

Another danger related to this kind of knowledge is to develop a relativistic approach to the idea of truth. Which happens when one extends this category to all landscape of human knowledge. 

%%%%%%%%%%%%%%%%%%%%%%%%%%%%%%%%%%%%%%%%%%%%%%%%%%%%%%%%%%%%%%%%%%%%%%%%%%%%%%
%%%%%%%%%%%%%%%%%%%%%%%%%%%%%%%%%%%%%%%%%%%%%%%%%%%%%%%%%%%%%%%%%%%%%%%%%%%%%%
\section{Descriptive area: raw knowledge}
\label{sec:Descriptive}

A large part of human knowledge is simply descriptive. The difference with the scientific knowledge is that such knowledge does no come with a theoretical framework. Thus, relatively to the scientific land, the descriptive land is the set of collections of raw informations.

For instance he map of a stars at a given time does not create any knowlege outside what is on the map or on the map itself. In this, it is very different from the Newtonian theory of gravity which does provide new knowledge. Indeed, given the relative positions of a satellite and forces that apply to it, one can forecasts the future positions of the satellite. Another trivial example is that you cannot use a map of London to find your path in Paris. But the science of map drawing can help you to draw a map of anywhere you will find yourself.

From a computability point of view, this area looks like database management more than computation of new knowledge based on data and a theoretical model leading to the computation of new results. In some sense it can be seen as a particular, finite, example of theoretical knowledge in which one has axioms (recorded information) accompagned with a not very powerful query language, meaning that inferences are very straightforward.

\subsection*{Data Science and spurious correlations}

A recent scientific trend, though, called data science, is trying to expand the descriptive area through the use of statistical methods. It is also presented under the designations “Artificial Intelligence” and/or “Machine Learning”. Initially the role of statistics is to sort the wheat from the chaff and to eliminate the noise in order to recover the signal from noisy measurements. It has been recently overused as a supermodel of knowledge building strategy, with some going as far as calling science dead. For instance C. Anderson states that the scientific method is superseded by the deluge of data in \cite{Anderson2008}.  

Of course there is the problem of spurrious correlations, eg see\cite{LongoCalude}. But beyond that critic, there is, underlying to these approaches based on data, the assumption that behind every correlation there is a reason waiting to be found. A “reason” is a theorical model allowing to describe the phenomenon more economically than the phenomenon itslef. Moreover, it also presupposes that all the variables that may have any influence whatsoever with this “reason” have been recorded in a specific data collection. Those two points are highly debattable. For once the existence of a reason cannot be presupposed. As have been shown with Turing and G\"odels work, it turns out that undecidability is unavoidable as soon as you have a theory powerful enough. Therefore, most of the phenomenon do not, given any theory, have a reason to begin with. Second, even for phenomenon for which it is possible to build a theory, it is always a possiblity that the peculiar set of data at disposal does not contain the right data: either because of precision issues or “hidden variables” issues (ie one has not precisely identify what are the relevant pohenomons). The risks liknked to a purely statistical approach to knowledge has been well studied.

\subsection*{Fat Tail Events}

There is another deep problem coming with the confusion between descriptive and scientific knowledge. It is the problem of fat tail events. The idea has been popularized by N.N. Taleb in \cite{Taleb2007}. Basically it is the fact that for some phenomenon, typically markets valuation, there are very rare occurrences of very large events that dwarf the rest of the distribution. A basic example is that in a single day of a market crash some banks may lose all of what they ever earned. From an epistemological point of view the problem associated with fat tail events is that it is not easy to separate the wheat from the chaff: a fat tail distribution may mimic a thin tail distribution until a very rare, but very catastrophic event, occurs. This is the problem of induction, well captured with the Russell's inductivist turkey allegory. Imagine a turkey that is fed every day by, observationally benevolent, humans. Every day reinforces the belief of the turkey that humans are good and will give it care and food ... until Thanksgiving Day.

There is, though, an area where descriptive knowledge can be useful to make predictions. For instance, if you look at height or weight distribution, you have thin tail distributions. Such kind of distribution allows you to make basics previsions: typically you can the expected weight of x persons given the population statistics and then build an appropriately calibrated elevator based on this expectation (you know that seven persons will appoximately weight 550kgs). As mentioned earlier, it is typical of what a database-type query  returns. It is an information that is not strictly recorded, bu, basically, it is possible to infer this information from first order logic: Codd's theorem being a famous example of this \cite{Codd1972}.

%%%%%%%%%%%%%%%%%%%%%%%%%%%%%%%%%%%%%%%%%%%%%%%%%%%%%%%%%%%%%%%%%%%%%%%%%%%%%%
%%%%%%%%%%%%%%%%%%%%%%%%%%%%%%%%%%%%%%%%%%%%%%%%%%%%%%%%%%%%%%%%%%%%%%%%%%%%%%
\section{Subjective area: Humanities and the fabric of reality}
\label{sec:Fiat}

From an epistemological point of view the problem of the humanities is that, in this area, it is not even clear what "reality" stands for. The typical examples are psychological: self-consciousness, pain, feelings/sensations (like the sensation of the blue of the sky, the odor of fresh bread etc.). On one hand it is very hard to deny their existence, try to help a suffering person by telling him/her that there is no such thing as pain for instance. On the other hand they have no physical measurable direct instantiations: seeing brain activity of a personn feeling pain is not seeing pain itself. This issue extends to inter-subjective realities like languages or institutions that emerge from the interaction between many people, but that are not natural phenomenon. Typically, the kind of phenomenon described as catallactic \cite{Hayek}. For instance the very idea of being part of a nation has no biological, physical, or even purely subjective (you may share this feeling with others but it cannot exists within a single person) counterpart. It is therefore a field of knowledge about phenomenon that do not exist outside our existence. 

\subsection*{You know more than you can tell}

Humanities is this body of knowledge that is not directly transmissible. For example, to describe what reality is denoted by the word pain, one has to rely on introspection and past events of his life. The meaning of pain is not communicated via words. One has to elaborate on shared experiences to communicate about it. It marks a stark contrast with relation to mathematics, where the meaning is carried by the text, and natural sciences, where the meaning relies on measurable phenomenon. Note that measurements are nothing more than standardised experiments. And, of course, the inert world doesn't learn or adapt, so experiments are reproducible. It is not necessarilly the case for living entities, especially humans because they learn and adapt. There is a kind of meta Heiseinberg principle at play: once a sociological or psychological study has been done and published it affects what it was about in the first place. It is even the standard procedure, one could even go as far as saying the aim, to political action: first a study is requested, then, supposedly based on science, politicians edict rules and regulations in order to act on the problem of interest. Both the study, the conclusions, and the regulations may affect the society and make the so-called conclusions of the initial study no longer valid. A classic (or should be classic) example is the famous marshmallow test \cite{McCrory2018}. More importantly, it resulted in the replication crisis as we know it \cite{Ioannidis2005}.

Therefore phenomenon belonging to humanities are, by definition, prone to misunderstandings. Indeed, there are neither internal (coherence) nor external (measurable reality) guards and thus no constraints on the discourse. There is not even an equivalent of the global context in physics under which it is presumed that phenomenon are stable across places, cultures and times. This is partly why one can engineers hoaxes like Sokal's hoax \cite{Sokal} and Sokal squared \cite{Grievance}. 

This absence of check and balance can be seen either for dynamic and static production of knowledge in the realm of humanities. . 

It is very hard in this area of knowledge to have models or theories that meet the standards of the natural sciences. Indeed, phenomenon studied in humanities are directly measured by, litteraly, what people say about it. It can be indirectly measured by trying to tie a phenomenon to a manifestation (but the two are in fact separate). 

The problem in this area is that theories do not have solid links to an external reality like in natural sciences. Therefore things can go south really quickly because there are no clear ways to refute a theory.   
\section{Scale-variant area: Where art and science meet together}
\label{sec:artsorscience}

There are body of knowledge where the object of study is a subtle mix between subjective and objective. Two examples come to mind: medicine, for which the debate on whether it is an art or a science continues today e.g. see \cite{Panda2006}, and economics. It turns out that such questions are linked to scale issues. 

\subsection*{What does "good health" even means?}

A question as simple as: "how do you define good health?" gets more and more tricky as soon as one starts thinking seriously about it. A definition from the dictionary \cite{goodhealth} is not very helpful:
\begin{quote}
a relative state in which one is able to function well physically, mentally, socially, and spiritually in order to express the full range of one’s unique potentialities within the environment in which one is living.
\end{quote}

Indeed the words "well", "full range", "unique potentialities" could be discussed upon on many pages. For instance: does "well" means {\em good enough}? or {\em better than the average} (average for who? You nowadays? You across time? The entire population? etc.)?  And if someone is doing fine in three of the four categories mentionned in this definition, but not fine (whatever that means) in the fourth one. Does that mean that this person is ok or not? A maybe more practical way to define "good health" is to see it negatively. Let us define "good health" as an absence of harm. Though it is only slightly more informational since the word "harm" is introduced without a proper definition.

This, not so trivial, question is of particular importance. Indeed, how can be defined a field, here medicine, for which one doesn't even know how to define at what this field aims for? From an epistemological perspective it is even worse than what was discussed in section \ref{sec:Fiat}, where the object of study is well defined, even if it is subjective.

Note that the same kind of question raises the same kind of problems when economics is considered: What is a "good economy"? Does only the output have to be measured? What about environmental issues? What about wealth repartition? The list of questions raised by any definition is almost infinite.

\subsection*{Science at the micro-level, art at the macro-level?}

Things get even stranger when both fields are considered at a micro-level. The more small scale problems are considered, the more those fields behave, with relation to knowledge, like scientific fields. In medicine the study of antibiotics and their effects on various bacteria is as scientific as one can imagine. Now, consider the epidemiological studies on whether or not to use antibiotics to treat some cases, things get less clear. One has to take into account the effects of the antibiotics (what is the effectivness of the molecule to cure the disease) but one has to balance them with other factors: are there severe secondary effect? is the disease treated by this antibiotics very dangerous (maybe the use of antibiotics will only have a marginal impact)? On top of that you can add epidemiological concerns: one may want to limit the use of antibiotics in order to lessen the spread of resistant bugs. Economical considerations can also be of interest: from the point of view of the society, is it worthy or not? Etc. Consider the example of nutrition. The issue on how to properly feed oneself is not a what can be called a new problem. Hippocrates put it this way “let food be your medicine” some millennias ago. Today's obesity and diabetes epidemics show that little effective progress has been made in this area of health. On the contrary vaccination campaignsi, for instance, are much more focused and have led to sizeable successes. Studies after studies we have a moving definition of what a healthy diet could be. This eluding "good diet" theory is a perfect allegory  of the elusive definition of "good health". 

In economics, some results are purely mathematical results: for instance Ricardo's famous example \cite{Ricardo1817} on comparative advantages is a textbook mathematical solution of an optimization problem. On one hand, one can discuss about the applicability of this result in the real world, but one cannot discuss the mathematical result on itself. On the other hand, when a very large-scope point of view is considered, things get less and less scientific. The century old controversy between Keynesian and Austrian economics shows that, whatever your opinion on this particular debate is, there is nothing like a "science is settled" feeling in this field. It is mainly due to different perceptions on what constitute a good economy.  

\subsection*{Statistics as philosopher's stone}

One way to circumvent the problems encountered by large scope issues for which it is very hard to design useful theories, is to heavily use statistics. Though, as the replication crisis \cite{Ioannidis2005} amply shows, it does not solve the issue at all. And indeed, statistics are more suited to adress static knowledge than being used as a general mechanism to infer medical or economical theories. 

The use of statistics in natural sciences is of a different nature: for instance, in Physics, statistical methods are used to estimate error intervals. Statistical methods are here used not as a generic theoretical framework but as a way to recover a signal by mitigating the noise emanating from real world imperfections (of measurement instruments, error in calibration etc.). This is the legitimate use of statistics.

The wider the scope of a medical phenomenon is, the more statistics are used as a veil for the lack of control on what variables matter relatively to the phenomenon of interest, and how actually causation works. The presupposition is not only that the future will be like the past, but also that what has been observed and measured is the only relevant part of the problem of interest.

It is bad practice both from both a theoretical and a practical point of view. From a theoretical point of view because everyone knows that correlations are not causations. It also not advisable because it promotes a lazy “one catch all” attitude towards the scientific method: one relies on a single meta model which is, in facts, pure statistics. From a practical point of view it can also be very bad: fat tail events are poster child phenomenon demonstrating the ineffectiveness of this approach in some situations. A typical example is the domain of finance: assessing assets return based on an incorrect use of statistics arguably led to market crashes that have impacted economy on a global scale \cite{Taleb2007}.

%%%%%%%%%%%%%%%%%%%%%%%%%%%%%%%%%%%%%%%%%%%%%%%%%%%%%%%%%%%%%%%%%%%%%%%%%%%%%
%%%%%%%%%%%%%%%%%%%%%%%%%%%%%%%%%%%%%%%%%%%%%%%%%%%%%%%%%%%%%%%%%%%%%%%%%%%%%%
\section{Conclusion}
\label{sec:Conclusion}

Computability theory plays a particular role in epistemology. Indeed, at its very core, computability theory is a mathematical study of structured knowledge. Ultimately, computability theroretical results have implications on many branches of knowledge, specially for the scientific ones in which objectivity is involved. Indeed, in order to have a useful theory one has to be able to actually compute in this theory. Using this point of view is interesting since it sheds a new light on different domains of knowledge acquisition techniques, and how they relate to one another. Knowing the limits of structured knowledge is useful and could be of help in many discussions. In the same way thermodynamics played a decisive role in stoping researches for perpertual motion, we believe that computability theory could play this role in other discussions.  

A particular point for which computability theory can be interesting lies in how unknowns are delt with. Contrary to standard approaches in which it is implicitly supposed that a particular piece of knowledge is either known or unknown (and in which double negation is understood as an affirmation), computability theory introduces these ideas around "semi-decidability" that are more subtle. Standard computability techniques, like the study of decidable subclasses of undecidable problems, may prove useful in this respect.  

Computability theory is a major intellectual feat of the XX$^0$-th century. It has not yet been largely used in discussions outside the scope of theoretical computer science. Though we think it is important that it is taken into account more often in other domains. In par with other great scientific ideas that help to shape the intellectual discussion, from relativity theory to quantum mechanics or genetics, computability theory notions may prove useful in many contexts.

%\begin{acknowledgements}
%\end{acknowledgements}

%%%%%%%%%%%%%%%%%%%%%%%%%%%%%%%%%%%%%%%%%%%%%%%%%%%%%%%%%%%%%%%%%%%%%%%%%%%%%%
%%%%%%%%%%%%%%%%%%%%%%%%%%%%%%%%%%%%%%%%%%%%%%%%%%%%%%%%%%%%%%%%%%%%%%%%%%%%%%
\bibliographystyle{alpha}
\bibliography{biblio}
\newcommand{\etalchar}[1]{$^{#1}$}

\end{document}